\documentclass[12pt]{article}
\usepackage{esfconf}
\usepackage{epsfig}
\usepackage{amssymb}
\begin{document}
\newcommand{\ncm}{\newcommand}
\ncm{\BDM}{\begin{displaymath}} \ncm{\EDM}{\end{displaymath}}
\ncm{\BEQ}{\begin{equation}}    \ncm{\EEQ}{\end{equation}}
\ncm{\BAR}{\begin{array}}       \ncm{\EAR}{\end{array}}
\ncm{\BEA}{\begin{eqnarray}}    \ncm{\EEA}{\end{eqnarray}}
\ncm{\Zm}{\mathbb{Z}}           \ncm{\WB}{\overline{W}}
\ncm{\om}{\omega}               \ncm{\sig}{\sigma}     \ncm{\al}{\alpha}
\ncm{\lam}{\lambda}             \ncm{\si}{\sigma}
\ncm{\Lb}{\left[}               \ncm{\Rb}{\right]}
\ncm{\lb}{\left\{}              \ncm{\rb}{\right\}}
\ncm{\lk}{\left(}               \ncm{\rk}{\right)}   
\ncm{\HH}{{\cal H}}             \ncm{\UU}{{\cal U}}    \ncm{\AAA}{{\cal A}} 
\ncm{\PP}{{\cal P}}             \ncm{\QQ}{{\cal Q}}
\ncm{\GG}{{\cal G}}             \ncm{\JJ}{{\cal J}}
\ncm{\hx}{\hspace*{2mm}}        \ncm{\hi}{\hspace*{12mm}}
\ncm{\hs}{\hspace*{1cm}}        \ncm{\su}{\sum_{j=1}^L}
\ncm{\hq}{\hspace*{6mm}}        \ncm{\ny}{\nonumber}
\ncm{\vns}{\vspace*{-2mm}}        
\ncm{\brv}{{\langle v|}}        \ncm{\ktv}{{|v\rangle}} 
\ncm{\ket}[1]{\left|\,{#1}\,\right\rangle}  \ncm{\ra}{\rightarrow}
\ncm{\bra}[1]{\left\langle\,{#1}\,\right|}  \ncm{\vph}{\varphi}
\ncm{\tfr}{\textstyle\frac}     \ncm{\hal}{{\tfr{1}{2}}} 
\ncm{\pih}{\tfr{\pi}{2}}        \ncm{\sui}{\vph=\phi=\pih} 
\title{Correlation functions and 
Corner Transfer Matrix of the Chiral Potts Model}
\authors{Chr. Ritter and G. von Gehlen}
\addresses{Physikalisches Institut der Universit\"at Bonn, Nussallee 12,
   53115 Bonn, Germany\\ 
   e-mail: christian.ritter@cii.de;\hspace{2mm} gehlen@th.physik.uni.bonn.de}
\maketitle
\footnotetext[1]{{\scriptsize To be published in "Quantization, Gauge Theory 
and Strings" dedicated to the memory of Efim Fradkin, {\it ed.} 
A.Semikhatov, M.Vasilliev and V.Zaikin, Moscow 2001.}}
\vspace*{-3mm}
\begin{abstract}
We use the Density Matrix Renormalization Group technique to compute 
correlation functions of the $\Zm_3$-chiral Potts quantum chain in the
massive regimes. Chains of up to 70 sites are used and a clear oscillatory 
behavior is found. We also check the relation between the density matrix
spectrum and the Corner Transfer Matrix spectrum pointed out by Nishino
and Okunishi, using Baxter's low-temperature formulae for the chiral
Potts Corner Transfer Matrix eigenvalues. We find very good agreement. 
\end{abstract}
\section{Introduction}\vspace*{-3mm}
We report numerical calculations which apply the density-matrix renormalization 
group method (DMRG) for the first time to the chiral Potts quantum chain. 
Although the chiral Potts model is integrable, various interesting
properties are not yet accessible to analytical calculation and one has to 
resort to numerical calculations in order to get more complete
information. We shall concentrate on two aspects: first, we show clearly 
that in the high-temperature massive phase the correlation functions are
oscillating. Second, we check the relation $\rho\sim ABCD$ between the
density matrix and the Corner Transfer Matrices (CTM), pointed out by
Nishino and Okunishi \cite{NiOk}. For the CTM eigenvalues of the chiral
Potts model we use Baxter's low-temperature expansion formulae
\cite{BCTM}. 
\vspace*{-3mm}
\section{The integrable chiral Potts model}\vspace*{-2mm}
The $\Zm_N$-symmetrical chiral Potts model has been introduced by Ostlund 
\cite{Ost} in order to describe incommensurate phase transitions in surface 
layers, e.g. Krypton monolayers on graphite. Centen {\it et al.}
\cite{Cen} introduced and analysed the chiral Potts
quantum chain. Very soon it was realized that this class of models can be
reformulated such that very interesting mathematical structures emerge:  
Howes, Kadanoff and den Nijs \cite{HKN} modified the $N=3$ quantum chain
to make it self-dual and discovered that in their version for a
particular chiral angle the lowest gap depends linearly on the inverse 
temperature. Then von Gehlen and Rittenberg \cite{GR} constructed a whole set of 
$\Zm_N$ symmetrical quantum chains which for any $N$ satisfy the Dolan-Grady 
\cite{DG} integrability conditions. These quantum chains are now usually called 
"superintegrable". Perk noticed \cite{Pk} that the Dolan-Grady conditions are 
just conditions for generating the algebra which Onsager in 1944 had introduced 
to obtain the free energy of the Ising model \cite{On}.\\ 
Onsager's 1944 technique for solving the Ising model made use of specific 
$\Zm_2$-features and it has not been possible to adapt it for solving the 
$\Zm_N$-model. The actual solution for the eigenvalues of the superintegrable 
model was achieved differently: In 1987 
Au-Yang {\it et al.} \cite{AuY} invented a two-dimensional $\Zm_N$-symmetrical 
Yang-Baxter-integrable statistical model, which contains the chiral Potts chain
hamiltonian as the derivative of its transfer matrix. Inversion relations and 
functional equations were found for this 2-dim. model and were used to obtain 
the ground state energy \cite{BxGr} and also analytic formulae for the 
excited levels \cite{AMCP,DKMC}. Hints that the chiral Potts model is
related to 
the six-vertex model at quantum group parameter at root of unity came from 
Kore\-panov \cite{Kor} and the connection was shown by Bazhanov and
Stroganov\cite{BS}. 
\par We first summarize some basic notions of the 2-dimensional integrable 
chiral 
Potts model of \cite{AuY}. Fig.1 shows part of a diagonal square lattice and 
rapidity lines used to define the Boltzmann weights
\begin{figure}[!ht]
\centerline{\epsfig{file=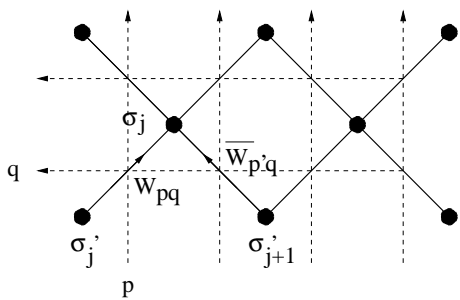,height=3.85cm,width=6cm,angle=0}}
\vspace*{-3mm}
\caption{Diagonally drawn square lattice with $\Zm_N$-spins
$\sig_j$ on the
lattice sites. Dashed lines: $p$- and $q$-rapidity lines.
Boltzmann weights $W_{p,q}\lk\sig_j-\sig'_j\rk$ act in SW-NE-direction, 
$\WB_{p,q}\lk\sig_j-{\sig'}_{j+1}\rk$ in SE-NW-direction.} 
 \end{figure}
$W_{p,q}(\sig_j-{\sig'}_j)$ and \mbox{$\WB_{p,q}(\sig_j-{\sig'}_{j+1})$} 
which depend on both rapidities via six 
functions $x_p,y_p,\mu_p,x_q,y_q,\mu_q$ (for brevi\-ty, we write $x(q)$
as $x_q$, etc.): $W_{p,q}(0)=\WB_{p,q}(0)=1$ and for $\;n=1,\ldots,N-1\;$ 
\BDM W_{p,q}(n)=\lk\frac{\mu_p}{\mu_q}\rk^n
       \prod_{j=1}^n\frac{y_q-\om^j x_p}{y_p-\om^j x_q}; 
\hq \WB_{p,q}(n)=\lk\mu_p\mu_q\rk^n \prod_{j=1}^n\frac{\om x_p-\om^j x_q}
 {y_q-\om^j y_p}  \EDM    where $\om=e^{2\pi i/N}$. 
Imposing $\Zm_N$-symmetry: $W_{p,q}(n+N)=W_{p,q}(n)$ (same for $\WB$), 
implies that $x_q$ and $y_q$ are restricted to high genus Fermat-like curves:
\BEQ k\:(x_q^Ny_q^N+1)=x_q^N+y_q^N;\hq k\,x_q^N=1-k'\mu_q^{-N};
         \hq k\,y_q^N=1-k'\mu_q^N   \label{bw}\EEQ 
(the same also for $x_p,\;y_p,\;\mu_p$). $k'^2=1-k^2$ are temperature-like
parameters, the low-temperature limit will be $k'\ra 0\,.$ 
For given $k$, 
up to discrete choices of roots, eqs.(\ref{bw}) leave only two of 
the six functions independent, e.g. $x_q$ and $x_p$. 
\\ For $N=2$ the chiral Potts model reduces to the Ising model
and the Boltzmann weights can be written in terms of Jacobi
elliptic functions  
\BDM x_q=\sqrt{k}\;\mbox{sn}(q,k);\hq
  y_q=\sqrt{k}\; \mbox{cn}(q,k)/\mbox{dn}(q,k),\hq 
     \mu_q=\sqrt{k'}/\mbox{dn}(q,k). \EDM 
$W$ and $\WB$ satisfy star-triangle-relations which involve the $p$- and 
$q$-variables {\it separately}, {\it not} as usual only through the 
difference $p-q$:\vns
\BDM \sum_{l=0}^{N-1}\WB_{q,r}(l'-l)W_{p,r}(l''\!-l)\WB_{p,q}(l-l''')\!=\!
  R_{pqr}W_{p,q}(l''-l')\WB_{p,r}(l'-l''')W_{q,r}(l''-l''').\vns  \EDM
$R_{pqr}$ can be written in terms of the $W$, $\WB$ and their $N$-th roots
\cite{AuY,MatSm}.
From the Boltzmann weights, the transfer matrix is defined by \vns\vns 
\BEQ    T_{p,q}(\{l\},\{l'\})\;=\;
\prod_{j=1}^L\;\WB_{p,q}(l_j-{l'}_j)\;W_{p,q}(l_j-{l'}_{j+1}).\label{tra}\vns\EEQ 
and the chiral Potts quantum chain hamiltonian arises as the derivative \vns
\BEQ \lim_{q\ra p}\frac{\partial T_{p,q}}{\partial q}\sim H
=-\sum_{j=1}^L\sum_{l=1}^{N-1}\lk \al_l\: Z_j^l                                                          
+\lam\:\bar{\al}_l\;X_{j}^l \,X_{j+1}^{N-l}\rk,\label{ham}\vns \EEQ  
where the operators $X_j$ and $Z_j$ act in the spaces ${\mathbb{C}}^N$ at sites 
$j$ and satisfy \BEQ Z_i X_j=X_j\:Z_i\:\om^{\delta_{i,j}};\hi Z_j^N= X_j^N=1;\hi
 \om=e^{2\pi i/N};  \EEQ
\BEQ  \!\,\mbox{and}\hs
\al_l=\frac{e^{i\vph(2l-N)/N}}{\sin{\frac{\pi l}{N}}}; \hs\hq 
\bar{\al}_l=\frac{e^{i\phi(2l-N)/N}}{\sin{\frac{\pi l}{N}}};\hs\hx\; \lam=1/k'.
  \hq\label{ca}\EEQ                           
In right hand side of (\ref{ham}) we skipped an additive constant term.
$H$ commutes with the $\Zm_N$-charge operator $\hat{Q}=\prod_{j=1}^L Z_j$.
The eigenvalues of $\hat{Q}$ are $\om^Q$ with $Q=0,\ldots,N-1$.
Since $T_{p,q}$ depends {\it not} only on $p-q$, it matters, at which $p$ the
limit $q\ra p$ is taken. So, for given $\lam$, the hamiltonian (\ref{ham}) depends
on one chiral parameter $\phi$. The coefficients (\ref{ca}) are conveniently 
written in terms of {\it two} chiral angles $\vph$ and $\phi$, which are 
related through \vspace*{-2mm} 
\BEQ \cos{\vph}\:=\:\lam\,\cos{\phi}. \label{inc}\vspace*{-3mm}\EEQ 
Using $\vph$ as the independent variable, $\phi$ becomes complex for 
$\lam<|\cos{\vph}|$ and the hamiltonian (\ref{ham}) becomes non-hermitian.
The degrees of freedom remain unchanged, and the previously 
parity-nonconserving $H$ now acquires a parity symmetry. Baxter mostly considers
the 2-d model in this parameter range with $\vph<\pih$, since there
the Boltzmann weights are positive real. In the single point $\lam=1$ and $\vph=0$
we recover the Fateev-Zamolodchikov parafermionic model \cite{FZ}, 
see e.g. \cite{AlFZ}.\\ Performing a duality transformation on (\ref{ham}) we 
reach a model of the same type, which has $\phi$ and $\vph$ 
interchanged and $\lam$ replaced by $\lam^{-1}$ but also different b.c. 
Howes {\it et al.} \cite{HKN} originally considered a 
self-dual model with $\vph=\phi$ instead of (\ref{inc}).
Probably this model is not integrable except at the line $\vph=\pih$ and the 
point $\vph=0,\;\lam=1,$ where it coincides with (\ref{inc}).
\par The analysis of the general 2-d model for $N\ge 3$ is difficult since  
for general $k$ no {\it simple} hyperelliptic parametrization of the $W_{p,q},
\;\WB_{p,q}$ is possible \cite{BxHyp}. Nevertheless, the free energy and the 
interface tension have been obtained circumventing an explicit uniformisation 
\cite{BxGr,ORB}. In \cite{MCR} a gap-formula for the quantum chain 
has been obtained, from which the phase structure can be calculated: In the 
massive phases the 
ground state of the chain hamiltonian is translational invariant (momentum
$P=0$). For the high-temperature (small $\lam$) massive phase the ground state 
is in the sector $Q=0$. In the low-temperature phase 
varying $\lam, \vph$ or $L$, the ground state rotates among the $P=0$, $Q$-sectors
with many level-crossings. 
The incommensurate (IC) phase is reached when a $P\neq 0$-state 
dips below the lowest $P=0$ state. The IC phase turns out to have a wedge shape,
the tip of the wedge is at $\lam=1,\;\vph=0$ (the parafermionic point) and 
spreading out with increasing $\vph$. At $\vph=\phi=\pih$ for $N=3$ the IC-phase
covers the interval $0.9013<\lam<1.1095$. 
\par Much more detailed analytic information has been obtained for the special 
"superintegrable" case $x_p=y_p,\;\mu_p=1$, or, in terms of the chain parameters:
 $\sui$. In this case $H$ simplifies to \vspace*{-2mm}
\BEQ H^{(s)}=-\:\frac{4}{N}\:\sum_{j=1}^L\sum_{l=1}^{N-1}\frac{1}{1-\om^{-l}}
 \lk Z_j^l\;+\:\lam\:X_j^l X_{j+1}^{N-l}\rk.\vspace*{-2mm}\EEQ 
Splitting $H^{(s)}$ into two pieces by writing 
$-H^{(s)}=A_0+\lam A_1$, one finds \cite{GR} that $\;A_0$ and $A_1$ 
satisfy the Dolan-Grady \cite{DG} conditions
\BEQ [A_0,[A_0,[A_0,A_1]]]=16\,[A_0,A_1];\hs
             [A_1,[A_1,[A_1,A_0]]]=16\,[A_1,A_0].  \EEQ 
Iff these conditions are met, starting from $A_0,\;A_1$ one can generate 
Onsager's algebra $\;\AAA\;\: $\cite{On}: 
\BDM   [A_l,A_m] = 4\,G_{l-m};\hq\!\! 
 [G_l,A_m]= 2\,A_{m+l}-2\,A_{m-l};\hq\!\![G_l,G_m]= 0;\hq\!\! 
 l,\:m\in \Zm. \EDM $\AAA$ implies an
infinite set of $\lam$-dependent commuting charges $\,[Q_m,\,Q_{m'}]=0\,$:  
\BEQ Q_m=\hal \lk A_m+A_{-m}+\lam(A_{m+1}+A_{-m+1})\rk \hx\hx
\mbox{among which}\hx\hx
  Q_0\sim H^{(s)}. \EEQ
As a consequence of $\AAA$ \cite{Dav} all 
eigenvalues $E$ of $H^{(s)}$ depend on $\lam$ as\vns
\BEQ E=\,a\,+b\,\lam\:+\,\sum_{j=1}^{m_E}4\,m_j
                 \sqrt{1+2\lam\cos{\theta_j}+\lam^2}\vns \EEQ
where $a,\;b,\;m_E$ are integers, and one finds $m_j=\pm 1/2$. 
$\AAA\:$ gives no information about which values of $a,\;b$ 
and $\theta_j$ appear. Only few facts are known about the representation 
theory of 
$\AAA$ \cite{Dav,Dat}.\\
Starting from functional relations \cite{AMCP,BS,BBP}, and solving them via
Bethe-ansatz-like equations \cite{AMCP,Tar}, the complete finite $L$ spectrum 
for the $\Zm_3$ superintegrable quantum chain has been given in
\cite{DKMC}. The general $\Zm_N$-case is discussed in \cite{Bifac}. All $\Zm_N$ 
conserving boundary conditions (b.c.) 
$X_{L+1}=\om^R X_1;\;\;R=0,1,\ldots,N\!-\!1$ can be treated.
Eigenstates of $H^{(s)}$ are not known for general $\lam$. There are  
high- and low-temperature expansion results \cite{Ho}.  
\section{Correlation functions}
To get more information about the phases of the $Z_N$-chiral Potts 
quantum chains we are interested in correlation functions.
We assume periodic b.c. $X_{L+1}=X_1$ and will only consider 
expectation values in the lowest translational invariant state, which we denote
by $\ket{v}$. This is the ground state in the non-IC regimes\footnote
{Albertini and McCoy \cite{AC} have studied correlation functions in the 
superintegrable IC region via finite-size corrections to the energy levels and
methods from conformal field theory.}. 
We concentrate on two-point functions of local operators 
$\Xi_j$, where $\Xi_j$ stands for either $Z_j$ or $X_j,$ or powers of these 
operators. We assume $\bra{v} v\:\rangle=1$ and define as usual 
\BEQ C_\Xi(r)=\bra{v}\Xi_{j+r}^+\:\Xi_j\ket{v}-
        \bra{v}\Xi_{j+r}^+\ket{v}\bra{v}\Xi_{j}\ket{v}.    \EEQ
Since $\ket{v}$ is translational invariant, $C_\Xi(r)$ will be independent of
$j$. We have $C_X(-r)=C_X(r)^*;\;\;C_Z(-r)=C_Z^*(r)=C_Z(r)$. Thus, although the 
chiral Potts model is not parity invariant, with periodic b.c. we can restrict 
ourselves to positive $r$ with $0<r<L/2$. 
According to the asymptotic behavior expected in the massive phases we will
fit our numerical data at large $r$ to the form
\BEQ C_\Xi(r)\;\approx\; a_\Xi\;e^{-r/\xi_\Xi}\;e^{2\pi ir/\Lambda_\Xi}\;
   +\;(1-a_\Xi)\delta_{r,0}.       \EEQ 
A finite $\Lambda_\Xi$ means oscillatory correlations. In the high-temperature
regime $C_Z$ is real and so it cannot show oscillations. $C_Z$ is real also
in 8th-order low-temperature expansion \cite{Ho}. So, we concentrate on $C_X(r)$
and $\Lambda$ will always mean $\Lambda_X$. 
Honecker \cite{Ho} gave a rigorous argument that $C_X$ must oscillate for 
$\phi=\pi$ (this is a curve at $\lam<1$ and $\vph>\pih$) because of the 
shifted parity symmetry present there. 
Then, by continuity, we expect such oscillations  
in the adjacent part of the high-temperature phase too. For $\Zm_3$ at $\phi=\pi$
one gets $|\Lambda|=6$, whereas from high-temperature expansion 
$\lim_{\lam\ra 0}\Lambda=6\pi/{\Re}e\vph$, so that on the superintegrable line
for $\lam\ra 0$ we should have $\Lambda\ra 12$ and $\Lambda=\infty$ at $\phi=0$. 
Observe that for small $\lam$ a small change of $\vph$ produces a strong change 
of $\phi$ and so there we expect a fast $\phi$-dependence of $\Lambda$.\\
Information on $\Lambda$ for $\lam\ra 1_+$ can be obtained from a finite-size
analysis of the pattern of ground-state level crossings. This way \cite{GG} 
concludes that in the thermodynamic limit $\Lambda\sim (\lam-1)^{-2/N}$ for 
$\lam\ra 1_+$ and $\phi=\pih$.  
\par
Numerical calculations of $C_X$ using Lanczos-type diagonalization of the 
sparce matrix $H$ have been performed in \cite{GH}, but since only chain lengths
$L\le 14$ or $r\le 7$ can be managed, $\Lambda$ (which, as we saw, should be
6...12 or larger) could only be crudely determined. Here now the DMRG-method
can do much better, and we will now turn to describe our new calculations. 
\section{The DMRG-method}
For many years, say from 1980-1992, numerical investigations of quantum chain
spectra mainly used Lanczos-type diagonalisation of finite chain hamiltonians.  
In this method, the full space of states is considered, restricted only by 
possible conservation laws as e.g. momentum, $\Zm_N$-charge or parity. By an 
iterated application of the sparse hamiltonian matrix one achieves excellent 
convergence for typically about 10-20 lowest energy levels. For a $\Zm_3$-model 
chains up to about $L=14$ sites (ca.$10^5\times 10^5$-matrices for periodic b.c.)
can be treated without much effort. The extrapolation to the thermodynamic 
limit $L\ra\infty$ has to be based on these few $L=2,\ldots, 14$ data.\\
While in this method each chain length $L$ is treated separately, the DMRG method 
invented by White \cite{Wh} starts with small chains and enlarges the chain 
step by step. This method has 
become standard in condensed matter physics and we will be brief in 
summarizing the main idea at the specific example of a $\Zm_3$-chain with
free b.c.:\\[2mm]
We start with a small chain $B_L$ of e.g. $L=5$ sites. Then $H$ is a $m\times m$
matrix, where $m=3^5=243$.  
Then we add one site and obtain a chain $B_{L+1}$ of $L+1$ sites. We don't want 
to keep all $3m$ states of $B_{L+1}$, but nevertheless want to get a good 
approximation for the ground state vector.\\  
In order to find out, which states of $B_{L+1}$ are most important for our goal
(how to renormalise the system), we double the system by adding a system 
of the same size as shown in the Figure:\vspace*{-6mm}   
 \BDM\vspace*{-3mm}\unitlength0.5mm
  \hspace{-4cm}\underbrace{ \begin{picture}(67.5,25)(0,-2)
        \put(20,5){$B_L$}            \put(110,5){$\tilde{B}_L$}
        \put(58,7.5){\circle*{7}}    \put(77,7.5){\circle*{7}}
        \put(0,0){\line(0,1){15}}    \put(0,0){\line(1,0){45}}
        \put(45,0){\line(0,1){15}}   \put(0,15){\line(1,0){45}}
        \put(45,7.5){\line(1,0){45}} \put(90,0){\line(0,1){15}}
        \put(90,0){\line(1,0){45}}   \put(135,0){\line(0,1){15}}
        \put(90,15){\line(1,0){45}}  \end{picture}}_{\hbox{$B_{L+1}$}} \EDM
This $2(L+1)$ sites system is called "superblock". We compute the 
$3^2\,m^2\approx 5\cdot 10^5$-dimensional ground-state vector $\ket{\Phi}$ of 
this superblock by a Lanczos algorithm: \BEQ\vspace*{-2mm}
 \ket{\Phi}=\sum_{i,j}\Phi_{i,j}\ket{i}_{B}\otimes\ket{j}_{\tilde{B}} 
 \label{dema} 
 \EEQ and form the {\it reduced} density matrix taking the partial 
 trace over the $\tilde{B}$-states:  
\BEQ \rho_{i,i'}=\sum_{j} \Phi_{i,j} \Phi^{\ast}_{i',j}\,. \label{rede}
\vspace*{-4mm} \EEQ
We diagonalise the {\it reduced} ($729\times 729$) density matrix $\rho_{i,i'}$ 
by some QL-algorithm and obtain the diagonal elements $\rho_i$. Only those
about $m'\approx 80$ states with the largest $\rho_i$ are now kept to describe
the renormalized matrix $\:B_{L+1}\:$.
This procedure is repeated until we reach chains of about $L=80$ sites, still
described by a $m'\times m'$ renormalised hamiltonian matrix.\\  
This method is most efficient for free b.c. If one wants to use periodic b.c.,
it is convenient to add a system repeated in the {\it same} orientation   
for not having two big adjacent blocks when joining the
outer ends. We have not mentioned the decomposition into parity 
eigenstates, which is useful, even if parity is not conserved as it is in our 
chiral Potts model. For calculating matrix elements of operators, one has to
keep track of the transformation of these operators into the new bases, which
requires considerable effort.   
\par The DMRG method works very well even up to $L=80...200$ sites 
keeping just about $m'\approx 80\ldots 150$ states. The reason for 
this is the exponentially decreasing spectrum of the density matrix, which, 
as first pointed out by Nishino and Okunishi \cite{NiOk}, can be understood from 
the close relation of the DMRG procedure to Baxter's variational method for 
calculating CTM eigenvalues \cite{Baxb}. We shall make use of this connection 
in the last section.\\[1mm]  
\begin{figure}[!ht]
\centerline{\epsfig{file=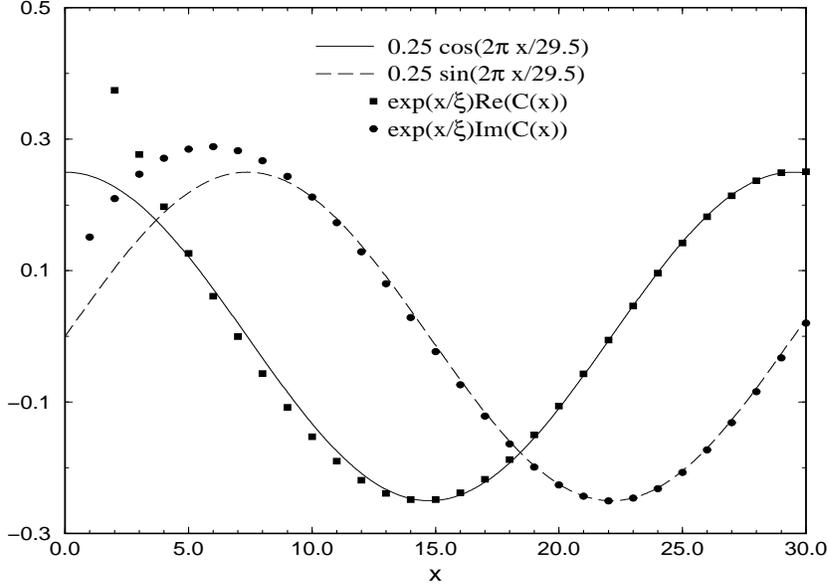,height=11cm,width=8cm,angle=-90}}
\caption{Real- and imaginary parts of the exponentially streched correlation
function $e^{x/\xi}C_X(x)$. Superintegrabel case $\sui$, high-temperature phase 
at $\lam=0.75$. The fit 
gives $\Lambda=29.5(5)$ and $\xi=2.000(5)$.}
\end{figure}
{\it \bf Results for the correlation functions}\\[1mm]
We have calculated $C_X(x)$ in the high-temperature phase from chains up to 
$L=80$ sites, so that it is reasonable to look up to $r=30$
sites. The so-called "infinite-size" DMRG approach \cite{Wh} is used, but
with periodic b.c. (although this is much more effort in DMRG) in order to
reduce boundary effects. Because of space limitations we give just one typical 
example for the superintegrable case and $\lam=0.75$. Of course, $C_X$ decreases
exponentially with $r$. So, in order to show the oscillation clearly, we multiply
$C_X$ by the factor $e^{r/\xi}$ which can be done since our numerical precision 
is quite high. We show both  the real and the imaginary part.\\[1mm] 
We fit also $\gamma$ in
$\xi^{-1}=2\,\gamma\,|1-\lam|$ (in the superintegrable case the gap is linear 
in $\lam$). Our fits give $\gamma=1.0\pm 0.1$ in lattice units. Going to 
smaller chiral angles (we give no Figures here, but these look similar to Fig.2)
we fitted the exponent $\nu$ in $\xi^{-1}\sim |1-\lam|^\nu$. $\nu$ is found to 
change little with $\vph$, not surprisingly since the mass-gap exponent 
$\nu=\frac{5}{6}$ for the parafermionic case $\phi=0$ is close to the value 
$\nu=1$ at $\phi=\pih$. However, $\Lambda$ changes 
strongly: it must diverge for $\phi\ra 0$. In the integrable 
case (\ref{inc}) the behavior of $\Lambda$ with decreasing $\phi$ is complicated 
because at $\lam<1$ the range in $\lam$ where the hamiltonian is hermitian, 
shrinks fast.
However, for the $\Zm_3$-self-dual (and probably not integrable) case $\phi=\vph$
within the errors we find $\Lambda\sim \phi^{-1}$.
\section{The Corner Transfer Matrix eigenvalues}
Baxter's corner transfer matrix (CTM) is a very efficient tool \cite{Baxb} for the 
calculation of single-spin expectation values (order parameters) of many 
integrable models. The CTM also plays also a crucial role in the Kyoto   
approach \cite{JM} to correlation functions of integrable models.\\[1mm] 
A CTM is defined as the partition function of a corner of a planar lattice 
(e.g. quadrant in case of a square lattice, or a sextant in case of a triangular 
lattice \cite{BCTM}). Fig. 3 shows a corner of a small square lattice. The
empty circles indicate the spins forming the indices of the CTM, full circles are
summed over, except for the lower diagonal spins, which are fixed or summed over, 
depending on whether we use fixed or free b.c. \vspace*{-2mm} 
\begin{figure}[!h]
\centerline{\epsfig{file=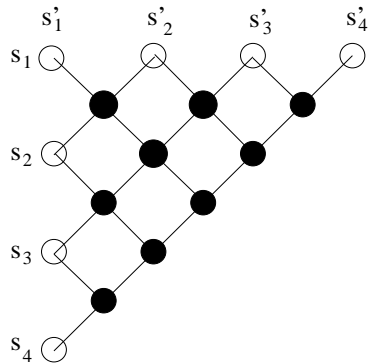,height=5cm,width=7cm,angle=0}}
\caption{Corner transfer matrix $A_{\{s\},\{s'\}}$}  \vspace*{-3mm}\end{figure}
The CTM of the SE-corner of a square lattice is usually called $A$, the other
anticlockwise rotated ones $B, C$ and $D$. The partition function is $Z=Tr(ABCD)$.
Baxter discovered that in the thermodynamic limit the spectrum of the CTM of the 
{\it eight-vertex model} takes a very simple form: let $A_d$ be $A$ diagonalised 
and normalized such that its maximum eigenvalue is unity. Then \vspace*{-3mm} 
     \BEQ {(A_d)}_{r,r}(u)=\exp{(-n_r x)}; \hs n_r\in \Zm, \label{Bct}
         \vspace*{-3mm}\EEQ
with $x=\pi u/K$ and $u$ is the rapidity parameter in the Boltzmann weights, 
$K$ the elliptic half-period.
For the derivation of (\ref{Bct}) the difference property of the Yang-Baxter 
equations is used and also some analyticity property which allows to make use 
of a periodicity property. From (\ref{Bct}) it is a short step to obtain the 
order parameter (magnetization).\\[1mm]  
Of course, also for the $\Zm_N$ chiral Potts model one is interested in the 
order parameters. We like to know $\langle Z^j_0\rangle$ for $j=1,\ldots,N-1$ 
where the spin with label 0 is far from the boundaries.
From series expansions and $Z$-invariance arguments there is the conjecture 
that the order parameters are independent of the chiral angle and their 
$k'$-dependence is \cite{ACPT}
 \BEQ \langle Z^j_0\rangle =(1-{k'}^2)^{j(N-j)/(2N^2)}.  \label{cj}\EEQ
Baxter \cite{BCTM,BFc98} has derived many important factorization and 
quasiperiodicity properties and functional relations of the chiral Potts CTM, 
but since the difference property is absent and analyticity seems to be 
restricted, as yet no exact proof of (\ref{cj}) was found. In the low-temperature 
limit Baxter did a truncated recursion calculation for the $\Zm_3$-case 
\cite{BCTM}. He obtains a 7th-order series approximation to the 17 largest 
eigenvalues $M_i$ of $M=ABCD$. These eigenvalues depend only on a 
temperature-like parameter, arising from the nome of the hyperelliptic 
functions which parametrize the Boltzmann weights. In detail: 
{\small \vspace*{-3mm} \BDM M_2\;=\;M_3\;=\;\tau;\hs 
 M_4=M_5=2\tau^3-\hal\tau^4-\frac{29}{8}\tau^5-\frac{5}{16}\tau^6
 +\frac{4147}{128}\tau^7+\ldots \EDM \BDM 
 M_{6\atop 9}=(3\pm 2\sqrt{2})\tau^4+(\mp 4\sqrt{2}-5)\tau^5-
 \lk \frac{5}{2}\pm\frac{49}{4\sqrt{2}}\rk\tau^6+\lk 12\pm\frac{377}{4\sqrt{2}}
  \rk\tau^7 +\ldots \EDM
 \BDM M_7=M_8=\hal\tau^4+\frac{13}{8}\tau^5+\frac{37}{16}\tau^6
   -\frac{2995}{128}\tau^7+\ldots \EDM
 \BDM M_{10}=M_{11}=3\tau^5+\frac{11}{3}\tau^6-\frac{730}{27}\tau^7+\ldots;\hq\;
 M_{13}=M_{14}=\frac{4}{3}\tau^6+\frac{29}{54}\tau^7+\ldots \EDM   \BEQ
 M_{12\atop 15}=(4\pm 2\sqrt{3})\tau^6+(10\pm 6\sqrt{3})\tau^7+\ldots;\hs
 M_{16}=M_{17}=\frac{9}{2}\tau^7+\ldots.\label{MDI} \EEQ}  $\!\!\!\!\!$
Here the largest eigenvalue is normalized to $M_1=1$ and $M_2$ is taken as the 
temperature scale $\tau$, which is related to $k'$ of (\ref{bw}) or rather,
$u={k'}^2/27$, by  \BEQ \tau=u+14u^2+254u^3+5150u^4+111123u^5+2495269u^6
 +57608712u^7+\ldots   \label{expa} \EEQ
In contrast to other integrable models with difference property, 
here the spectrum of $\log{M}$ is {\it not} equidistant. Baxter uses fixed 
spin zero b.c. on the CTM, so the ground state is non-degenerate. The 
degenerate levels are those with center spin 1 and 2, the non-degenerate ones
have center spin zero.
Baxter's derivation of (\ref{MDI}) is best justified in the region of real 
positive Boltzmann weights 
and it is not clear which details can be trusted when extending the formulae to 
the region of complex Boltzmann weights. Using (\ref{MDI}), (\ref{expa}) one 
finds correctly 
$\langle Z_0 \rangle={k'}^{2/9}$ up to the order $\tau^7$ considered, in 
agreement 
with (\ref{cj}) as first conjectured from the superintegrable chain. 
\\ Looking into the apparent convergence of the expansions (\ref{MDI}), 
(\ref{expa}), one is inclined to trust the lowest eigenvalues $M_2 \ldots M_5$ 
up to about $k'\approx 0.7$ within a few percent: For
$k'=0.7$ one finds $M_2\equiv \tau\approx 0.025$. However, approaching $k'=1$   
we clearly see that convergence of the series gets lost quite suddenly,
although in 7th-order at $k'=0.9$ we still remain at $\tau\approx 0.06$. 
\section{CTM eigenvalues from DMRG density matrices}
Fig.4 illustrates that the matrix $M=ABCD$ is just the partition function of 
a lattice with a seam opened between $A$ and $D$.
\begin{figure}[h]
\centerline{\epsfig{file=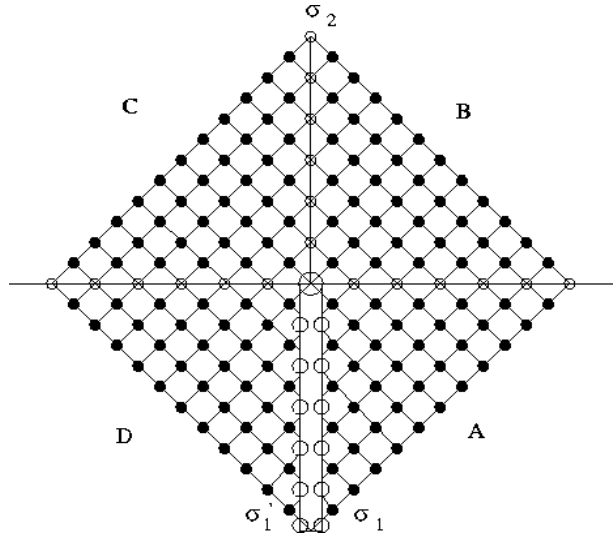,height=7cm,width=8cm,angle=0}}
\caption{The product of four corner transfer matrixes 
$(ABCD)_{\{\sig\},\{\sig'\}}$. The seam in the lower vertical indicates 
that no trace is taken.}   \end{figure}
There is a direct relation between $M$ and
the reduced density matrix $\rho_{i,i'}$ of (\ref{rede}): Baxter \cite{Baxb}, 
chapt.13, pointed out that in the thermodynamic limit the product of {\it two} 
CTM, say $AB$ in Fig.4,  
is the maximal eigenvector $\ket{\Psi}$ of the column-to-column transfer matrix 
(TM) (here moving left to right). Now, for 
integrable models the TM of the 2-dim. system is in one-to-one
correspondence with the quantum chain hamiltonian $H$ as e.g. in (\ref{ham}) 
(vice-versa one uses the Trotter construction)
and the maximal eigenvector of the TM is the ground state eigenvector of
$H$, i.e. $\ket{\Psi}\sim \ket{\Phi}$ of eq.(\ref{dema}). So $M=ABCD$, which is
summed over the closed seam betwen $B$ and $C$, 
is proportional to $\rho_{i,i'}$ of eq.(\ref{rede}). All this is correct 
for large systems, i.e. when the boundaries
play no essential role.\\[2mm] 
We now compare numerically the $\rho_{i,i'}$ (which appeared as a technical tool 
for selecting the most relevant states in our DMRG calculation) to Baxter's CTM 
eigenvalues (\ref{MDI}). Baxter uses fixed b.c. at low $k'$. It is not easy to 
implement fixed b.c. in DMRG, so we use free b.c. for the {\it dual}
model \cite{BxDu} at high temperatures $\lam<1$. We have studied several 
values of $L$ and $\vph,\;\lam$. In Fig.5 we show that there is excellent 
agreement already when we use a small $L=10$ system, which can be understood
considering that for $\lam\le 0.75$ the correlation length $\xi$ is smaller 
than three lattice sites. E.g. for $\lam=0.5$ eqs.(\ref{MDI}) give $M_4=
2.44884\cdot 10^{-6}$ whereas the diagonalisation of the density matrix
gives $\rho_4=2.4418\cdot 10^{-6}$ at 
$\phi=\pih$ and $\rho_4=2.4382\cdot 10^{-6}$ at $\phi=0$.
Baxter's formulae are not valid at $\lam=1$, so any agreement there is not 
significant. For the Ising and XXZ-models, the checks analogous to the ones
given here were made earlier by Peschel {\it et al.} \cite{Pes}. 
In their cases the CTM-eigenvalues are known exactly and, apart from 
degeneracies, are strictly logarithmically equidistant.\\[3mm]
G.v.G. gratefully acknowledges the support by INTAS-97-1312.\\[1mm] 
\vspace*{-3mm}
\begin{figure}[t]
\centerline{\epsfig{file=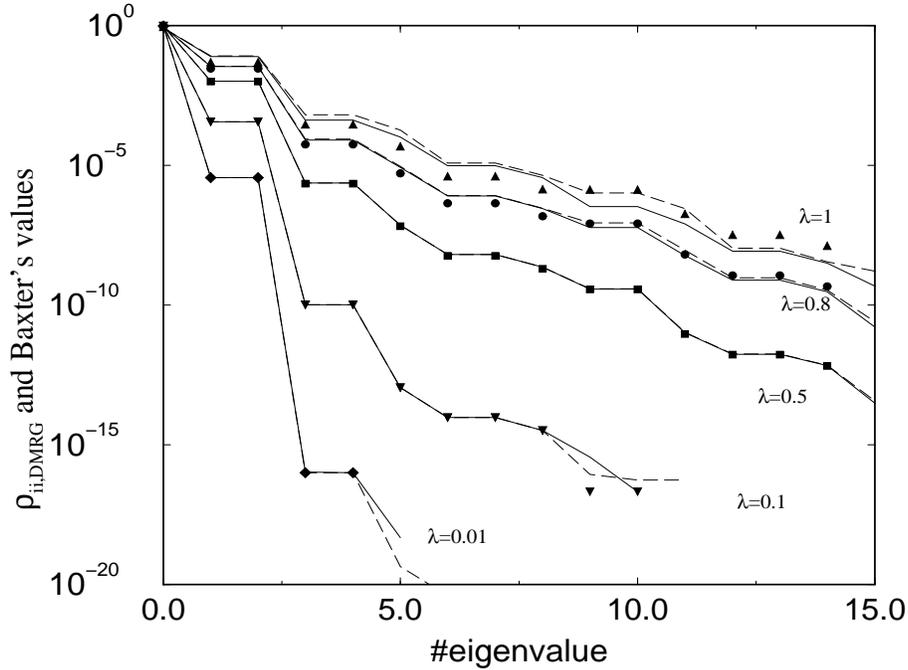,height=12cm,width=9cm,angle=-90}}
\vspace*{-3mm}
\caption{The largest 15 eigenvalues for several $\lam$. Discrete
points: Baxter's 
CTM-calculation (\ref{MDI}) together with our numerical 
density matrix eigenvalues for free b.c. and $L=10$: 
solid lines: at the boundary of the hermitian region $\phi=0,\;\vph=
\arccos{\lam}$; dashed lines: on the superintegrable line.\vspace*{-3mm}}  
\end{figure} 
\end{document}